\documentclass{article}
\usepackage{epsfig}
\usepackage{graphicx}
\usepackage{graphics}

\def\a{\alpha}

\def\t{\theta}
\def\bt{{\bar{\theta}}}
\def\yb{{\bar{y}}}
\def\Db{{\bar{D}}}
\def\Tr{{\rm Tr}}
\def\dis{\displaystyle}
\def\le{\left(}
\def\ri{\right)}
\def\da{{\dot{\alpha}}}
\def\no{\nonumber}
\def\del{\delta}
\def\bW{\bar{W}}
\def\G{\Gamma}
\def\rar{\rightarrow}
\def\fra1g2{\frac{1}{g^2}}
\def\dg{{\dagger}}
\def\Vt{\tilde{V}}
\def\Kt{\tilde{K}}

\def\e{\epsilon}

\def\f12{\frac{1}{2}}
\def\F{\Phi}

\def\F{\Phi}
\def\bF{\bar{\Phi}}

\begin{document}
\begin{titlepage}
\flushright{USM-TH-161}\\
\vskip 2cm               
\begin{center}
{\Large \bf Finiteness of ${\cal N} =4$ super-Yang--Mills effective action \\
\vskip 3mm
in terms of dressed  ${\cal N} =1$ superfields} \\
\vskip 1cm  
Igor Kondrashuk$^{a,b}$ and Ivan Schmidt$^a$ \\
\vskip 5mm  
{\it  (a) Departamento de F\'\i sica, Universidad T\'ecnica 
Federico Santa Mar\'\i a, \\
 Avenida Espa\~{n}a 1680, Casilla 110-V, Valparaiso 2390123, Chile} \\
\vskip 2mm 
{\it  (b) Grupo de Matem\'atica Aplicada {\rm \&} Grupo de F\'isica de Altas Energ\'ias  {\rm \&}  Centro de Ciencias Exactas   {\rm \&}  Departamento de Ciencias B\'asicas, Universidad del B\'io-B\'io, Campus Fernando May, Av. Andres Bello 720, Casilla 447, Chill\'an 3780227, Chile} \\
\end{center}

\begin{abstract}
We argue in favor of the independence on any scale, ultraviolet or infrared, in  kernels 
of the effective action expressed in terms of  dressed  
${\cal N} =1$ superfields for the case of ${\cal N} =4$ super-Yang--Mills 
theory. Under  ``scale independence '' of the effective action of dressed mean superfields 
we mean its `` finiteness in the off-shell limit of removing all the regularizations.'' This off-shell limit is scale independent because no scale remains inside these kernels after removing the regularizations. We use two types of regularization:  regularization by dimensional reduction and 
regularization by higher derivatives in its supersymmetric form. Based on the Slavnov--Taylor 
identity we show that dressed fields of matter and of vector multiplets can be introduced to 
express the effective action in terms of them. Kernels of the effective action 
expressed in terms of such dressed effective fields do not depend on the ultraviolet scale.
In the case of  dimensional reduction, by using the 
developed technique we show how the problem of inconsistency of the dimensional reduction 
can be solved. Using Piguet and Sibold formalism, we indicate that the dependence on 
the infrared scale disappears off shell in both the regularizations. 
\end{abstract}
\end{titlepage}

\section{Introduction}

The effective action is restricted by consequences of various symmetries of the 
classical action that at the quantum level take the form of specific identities.
One of them is the Slavnov--Taylor (ST) identity 
\cite{Slavnov:1972fg}-\cite{Zinn-Justin:1974mc}.   
This  generalizes the Ward--Takahashi identity of quantum electrodynamics 
to the non-Abelian case and can be derived starting from the property 
of invariance of the tree-level action with respect to BRST symmetry 
\cite{Becchi:1974md,Tyutin:1975qk}. The ST identity can be formulated  as 
equations involving  variational derivatives of the effective action. 
In the general ${\cal N} =1$  supersymmetric theory it can be 
written as \cite{Piguet:1996ys} 
\begin{eqnarray*}
& \Tr\left[\dis{\int~d^8z~\frac{\del \G}{\del V}\frac{\del \G}{\del K}
 - i \int~d^6y~\frac{\del \G}{\del c}\frac{\del \G}{\del L}
+ i \int~d^6\yb~\frac{\del \G}{\del \bar{c}}\frac{\del \G}{\del
\bar{L}}}\right. \no \\
& - \left. \dis{ \int~d^6y~\frac{\del \G}{\del b}\le\frac{1}{32}
\frac{1}{\a}\Db^2D^2V\ri
- \int~d^6\yb~\frac{\del \G}{\del
\bar{b}}\le\frac{1}{32}\frac{1}{\a}D^2\Db^2V\ri} \right] \\ 
& - \dis{i~\int~~d^6y~\frac{\del \G}{\del \F}~\frac{\del \G}{\del k} +
i~\int~d^6\yb~\frac{\del \G}{\del \bar{k}}~\frac{\del \G}{\del
\bF} = 0}. \no
\end{eqnarray*}
Here the standard definition of the measures in superspace is used 
\begin{eqnarray*}
d^8z \equiv d^4x~d^2\t~d^2\bt,~~~ d^6y \equiv d^4y~d^2\t,~~~
d^6\yb \equiv d^4\yb~d^2\bt.
\end{eqnarray*}
The effective action $\G$ generates one particle irreducible amplitudes of the quantum fields and contains all the information 
about the quantum behaviour of the theory. It is a functional of the effective fields $V,b,\bar{b},c,\bar{c},\F,\bF$ and 
external sources  $K,L,\bar{L},k,\bar{k},$ coupled at tree level to BRST-transformations of the corresponding classical 
fields \cite{Faddeev:1980be}. We use two types of  UV
regularization: regularization by higher derivatives \cite{Krivoshchekov:1978xg,West:1985jx} 
and regularization by dimensional reduction \cite{Siegel:1979wq,Capper:1979ns}.

We will show that the actual variables of the effective action are dressed 
effective superfields, that is, they are effective superfields convoluted 
with some unspecified dressing functions that are parts of propagators. 
Kernels accompanying  the dressed effective fields in the effective action are related to the 
scattering amplitudes of the particles. A similar problem has been solved in 
component formalism \cite{Cvetic:2004kx,Cvetic:2006kk}. As has been argued in Refs. \cite{Cvetic:2004kx,Cvetic:2006kk} 
in that formalism the dressed mean fields appear to be the actual variables of the effective action,
leaving the kernels of the action independent of any scale in the limit of removing all the regularizations, ultraviolet or off-shell infrared, in case of   ${\cal N} =4$ supersymmetric 
Yang--Mills theory. This statement has been confirmed by the explicit calculation 
in Ref. \cite{Cvetic:2006iu}. In the present paper this problem is considered by using ${\cal N} =1$ superfield formalism
which keeps one of the supersymmetries apparent and conserves all the R-symmetries which we need to apply anomaly multiplet 
ideas \cite{Sohnius:1981sn}.  The important point of Refs. 
\cite{Cvetic:2004kx,Cvetic:2006kk,Cvetic:2002dx,Cvetic:2002in,Kondrashuk:2000br,Kang:2004cs,Kondrashuk:2003tw,Kondrashuk:2001yd} 
is the possibility to absorb the two point proper functions in the re-definition of the effective fields.

The on-shell IR divergences are constrained in QFT by the well-known Kinoshita theorem, while they cancel in all observables \cite{Kinoshita:1962ur}.  However, ${\cal N} =4$ SYM is not a physical theory, it does not have any observables. How the finiteness of the kernels of this effective action  in the limit of the removing regularization is combined with this Kinoshita theorem?   Suppose we removed the regularizations in these proper correlators of dressed mean superfields. These kernels become scale-independent after removing these regularizations.   The corresponding connected Green functions (connected correlators) of the dressed mean fields contribute to the amplitudes on the mass shell. The amplitudes which are roughly on-shell values of these connected correlators may have IR divergences after putting them on-shell.  In order to work with the connected correlators on shell we need a regularization again. This may be any regularization which makes these on-shell values non-singular. These IR divergences may be regularized in a way which has nothing to do with the previous two regularizations,
ultraviolet and infrared, used initially for the UV renormalization procedure and for the suppression of the off-shell IR infinities of the ${\cal N} =1$ SUSY gauge theory. The most convenient way from our point of view may be  via MB parametrization. We proposed it   in Section \ref{sec:three}. It is important that this auxiliary regularization may be removed then at the end in the results for ``physical quantities''  
in the ``physical processes'' which are  ${\cal N} =4$ SUSY models of DIS structure functions
\cite{Alvarez:2016juq,Bianchi:2013sta,Alvarez:2019eaa,Kondrashuk:2019cwi}.

These two words ``finiteness'' and ``scale-independence'' are frequently used concepts in the framework for the AdS-CFT correspondence, where the correlators of the BPS gauge invariant operators may be found or restricted by using the methods of the conformal field theory. In this sense, ${\cal N} =4$ SYM is exactly solvable QFT. The insertions of composite non-BPS operators as external states into the correlators of the components of 
${\cal N} =1$ supermultiplets are treated as the only source of 
UV-divergences \cite{Minahan:2002ve}. In contrary, in the present paper  we talk about the finiteness (and, then as a result of finiteness, about the scale independence) of the connected correlators (they are built from the proper correlators) which contribute to scattering amplitudes in this theory in the Landau gauge 
\cite{Cvetic:2004kx,Cvetic:2006kk}.  Renormalization by the parts of the propagators has been proposed in \cite{Kondrashuk:2000br,Cvetic:2002dx}.

\section{ST identity in ${\cal N} =1$ SUSY formalism for ${\cal N} =4$ SYM}

We consider the {${\cal N}=4$ theory in {${\cal N}=1$ superfield 
formalism. This model has specific field contents. The Lagrangian of the model in terms of  ${\cal N}=1$ 
superfields is 
\begin{eqnarray*}
& S = \dis{\fra1g2\frac{1}{128}\Tr~\left[\int~d^6y~W_\a W^\a  +
\int~d^6\yb~\bW^\da \bW_\da \right.} \nonumber\\
& + \dis{\left.\int~d^8z~e^{-V}\bF_i~e^V~\F^i \right.}  \\
& + \dis{\left. \frac{1}{3!}\int~d^6y~i\e_{ijk}\F^i[\F^j,\F^k] 
+ \frac{1}{3!}\int~d^6\yb~i\e^{ijk}\bF_i[\bF_j,\bF_k]\right]}.
\end{eqnarray*}
For the {${\cal N}=1$ supersymmetry we use the notation of Ref. 
\cite{Kondrashuk:2000qb}. This Lagrangian 
for {${\cal N}=4$ supersymmetry is taken from  Ref. \cite{Gates:1983nr}. 
The flavor indices of the matter run in  
$i =1,2,3$ and the matter superfields are in the adjoint representation of the 
gauge group, $\F^i = \F^{ia}~T^a.$ (The vector superfield is in the same 
representation, $V = V^aT^a$).

Consider for the beginning the general ${\cal N} = 1$ super-Yang--Mills whose 
classical action takes the form
\begin{eqnarray}
& S = \dis{\int~d^6y~\frac{1}{128}\fra1g2\Tr~ W_\a W^\a  +
\int~d^6\yb~\frac{1}{128}\fra1g2\Tr~ \bW^\da \bW_\da } \nonumber\\
& + \dis{\int~d^8z~\bF~e^V~\F}  \label{mattact}\\
& + \dis{\int~d^6y~\left[Y^{ijk}\F_i\F_j\F_k + M^{ij}\F_i\F_j \right]}
+ \dis{\int~d^6\yb~\left[\bar{Y}_{ijk}\bF^i\bF^j\bF^k \no
+ \bar{M}_{ij}\bF^i\bF^j \right]}.
\end{eqnarray}
We do not specify the representation of the matter fields here. It is some 
general reducible representation of the gauge group with a set of irreducible
representations. The Yukawa couplings $Y^{ijk}$ and  $M^{ij}$
appear at some general triple vertex and mass terms in four dimensions. 
The path integral describing the quantum theory is defined as
\begin{eqnarray}
& \dis{Z[J,\eta,\bar{\eta},\rho,\bar{\rho},j,\bar{j},K,L,\bar{L},k,\bar{k}] 
= \int~
dV~dc~d\bar{c}~db~d\bar{b}~d\F~d\bF~\exp i}\left[\dis{S}
\right.  \no \\
& \left. + \dis{2~\Tr\le\int~d^8z~JV + i\int~d^6y~\eta c
+ i\int~d^6\yb~\bar{\eta}\bar{c} + i\int~d^6y~\rho b
+ i\int~d^6\yb~\bar{\rho}\bar{b}\ri} \right. \no\\
& \left. + \dis{\le\int~d^6y~\F~j + \int~d^6\yb~\bar{j}~\bF\ri }
    \right. \label{pathRM}\\
& \left. + \dis{2~\Tr\le i\int~d^8z~K\del_{\bar{c},c}V + \int~d^6y~Lc^2 +
\int~d^6\yb~\bar{L}\bar{c}^2 \ri }\right. \no\\
& \left. + \dis{\int~d^6y~k~c~\F + \int~d^6\yb~\bF~\bar{c}~\bar{k}} \right], 
\no
\end{eqnarray}

The derivation of the ST identity in general supersymmetric theory can be 
found for example in Ref. \cite{Kondrashuk:2001yd}. The result is 
\begin{eqnarray}
& \Tr\left[\dis{\int~d^8z~\frac{\del \G}{\del V}\frac{\del \G}{\del K}
 - i \int~d^6y~\frac{\del \G}{\del c}\frac{\del \G}{\del L}
+ i \int~d^6\yb~\frac{\del \G}{\del \bar{c}}\frac{\del \G}{\del
\bar{L}}}\right. \no \\
& - \left. \dis{ \int~d^6y~\frac{\del \G}{\del b}\le\frac{1}{32}\frac{1}{\a}\Db^2D^2V\ri
- \int~d^6\yb~\frac{\del \G}{\del
\bar{b}}\le\frac{1}{32}\frac{1}{\a}D^2\Db^2V\ri} \right] \label{STrM}\\ 
& - \dis{i~\int~~d^6y~\frac{\del \G}{\del \F}~\frac{\del \G}{\del k} +
i~\int~d^6\yb~\frac{\del \G}{\del \bar{k}}~\frac{\del \G}{\del
\bF} = 0}. \no
\end{eqnarray}

Regularization is necessary to analyze the identities. First, we consider the regularization by dimensional reduction. 
Under this regularization the algebra of Lorentz indices is done in four dimensions but integration is done in $4-2\e$ dimensions
in the momentum or in the position space.  As  has been shown in Refs. \cite{Cvetic:2004kx,Cvetic:2006kk}, such a 
regularization is self-consistent at least in  
${\cal N}=4$ supersymmetric Yang--Mills theory in component formalism at all orders of the perturbation theory. In the next  
paragraphs we will show such a regularization procedure can be applied at {\it all} orders of the perturbation theory
in superfield formalism too if this regularization is combined with Piguet and Sibold formalism of 
Refs. \cite{Piguet:1984mv,Piguet:1984im}

\section{From double-ghost $Lcc$ correlator to other correlators}  \label{sec:from}

Let's  analyse this identity in the following way. One can start by considering the monomial $Lcc$ of the effective action. 
Due to supersymmetry, superficial divergences are absent in chiral vertices \cite{Grisaru:1979wc,West:1986wu}. This theorem 
is a direct consequence of the Grassmannian integration and has been described, e.g., in Ref. \cite{Grisaru:1979wc}.  
However, there could be finite contributions.  They remain finite in the limit of 
removing the ultraviolet regularization. For example, at one loop level one can find among others the following kernel structure 
for the correlator $Lcc$  \cite{Cvetic:2004kx}:
\begin{eqnarray*}
& \dis{\int~d^4\t~d^4x_1d^4x_2d^4x_3~\frac{1}{(x_1-x_2)^2(x_1-x_3)^2(x_2-x_3)^4}f^{bca}}\times \\
& \dis{ \times \le D^2~L^{a}(x_1,\t,\bt)\ri c^{b}(x_2,\t,\bt)c^{c}(x_3,\t,\bt).}
\end{eqnarray*}
An efficient form to parameterize this contribution via MB transformation is given in Section \ref{sec:three}.

Landau gauge is the specific case in gauge theories because we do not need to renormalize the gauge 
fixing parameter. Absence of the gauge parameter is enough condition to avoid this possible source of  appearance of 
the scale dependence in kernels of dressed mean superfields through the renormalization of the gauge parameter. From the effective 
action $\G$ we can extract the two point ghost proper correlator  $G(z-z'),$
\begin{eqnarray*}
G^{\dg} = G,
\end{eqnarray*}
and a two point connected ghost correlator $G^{-1}(z -z'),$ which is related to the previous one in the following way:
\begin{eqnarray*}
\int d^8z'~G(z_1 - z')~G^{-1}(z_2 - z') = \del^{(8)}(z_2 - z_1).
\end{eqnarray*}
This definition is valid in each order of perturbation theory. One can absorb this two point proper function into a non-local 
redefinition of the effective fields $K$ and $V$ in the following manner \cite{Kondrashuk:2000br}: 
\begin{eqnarray}
& \dis{\Vt \equiv  \int~d^8z'~V(z')~G^{-1}(z-z')}, \label{conv}\\
& \dis{\Kt \equiv \int~d^8z'~K(z')~G(z-z').} \no
\end{eqnarray}
One can see that the part of the ST identity without the gauge fixing term is covariant with respect to such a re-definition of 
the effective fields.  We will call the construction  (\ref{conv}) dressed effective (or mean) superfields. Proceeding at one 
loop level in terms of the dressed effective superfields, one can see from the ST identity that the divergence of the 
$\Kt c\Vt$ vertex must be canceled by the divergence of the $Lcc$ vertex. However, the ${\cal N} =1$ $Lcc$ vertex does 
not diverge at one loop 
level in momentum space due to supersymmetry.\footnote{For simplicity we consider the Landau gauge.} 
This means that the ST identity clearly 
shows that the  $\Kt c\Vt$ is also finite \cite{Cvetic:2004kx,Cvetic:2006kk}, that is, it does not diverge in the limit of 
removing the regularization. The rest of the UV divergence in the propagator of the dressed gauge fields $\Vt\Vt$ can be 
removed by redefining of the gauge coupling constant. In theories with zero beta function, in case of this paper it is ${\cal N}=4$ 
supersymmetric Yang-Mills theory, this last divergence is absent. The other graphs are finite since proper correlators can be 
constructed from the $\Kt c\Vt$ (or $Lcc$), and $\Vt\Vt$ correlators by means of the ST identity. 
This is a direct consequence of the ST identity and $R$-operation \cite{Cvetic:2004kx,Bogolyubov:1980nc}. 
We can repeat this argument in each order of perturbation theory, which is  completely the same as in 
Refs. \cite{Cvetic:2004kx,Cvetic:2006kk} in component formalism. 

Why  we may copy this proof from the component formalism of Refs.  \cite{Cvetic:2004kx,Cvetic:2006kk}? This proof described in the previous paragraph is based on the structure of the gauge part and on the BRST symmetry of the classical action extended by additional BRST-invariant monomials \cite{Faddeev:1980be}. The BRST symmetry   is the origin of the ST identity \cite{Faddeev:1980be,Kondrashuk:2000br,Cvetic:2002dx}.
This is why the proof  in the previous paragraph may be copied from the nonsupersymmetric case. The only difference is that in the Wess-Zumino gauge 
the components of the same multiplet are treated separately in this proof.  Their renormalization constants do not coincide (this may be seen for example in Chern-Simons theory Ref.\cite{Avdeev:1992jt}),
however it does not present any obstacle. The superficial divergence of the $Lcc$ vertex  
is absent in both the formalisms (superfield and component) but by the different reasons. In the component formalism it happens due to the transversality of the Landau gauge which does not allow to this double-ghost vertex to diverge superficially  \cite{Cvetic:2004kx,Cvetic:2006kk}.

Thus, all the gauge part of the effective action can be 
considered as the functional of the effective fields $\Vt,$  $\Kt,$ $L,$ $c,$
$\bar{c},$ $\tilde{b}$ and $\tilde{\bar{b}}.$  Due to the antighost equation 
\cite{Kondrashuk:2000br}, the antighost field $b$ is always 
dressed in the same manner as the auxiliary field $K$ is dressed. Kernels of 
this effective action are functions of the gauge coupling, mutual distances and 
in general of the ultraviolet scale, because the divergences in subgraphs must be 
removed by the renormalization of the gauge coupling. But if the beta function is 
zero the kernels have no UV scale dependence. In the position space in component formalism the infrared divergences can be 
analysed in the same way like ultraviolet divergences were analysed in Ref.\cite{Bogolyubov:1980nc} in  momentum space 
by means of $R-$operation \cite{Cvetic:2006iu}. However, it is difficult to repeat this argument in the position space 
in superfield formalism since the propagator of vector superfield is dangerous in the infrared region. 
In view of this difficulty we use the formalism developed in Refs.  \cite{Piguet:1984mv,Piguet:1984im}, where 
the problem of off-shell infrared divergences of superfield formalism has been solved.

\section{Off-shell IR divergence in superfield formalism}

The infrared regulator has been introduced by means of the following trick of renormalization of the vector gauge field $V:$
\begin{eqnarray*}
V \rar V + \mu {\t^2}{\bt^2}V,
\end{eqnarray*} 
where $\mu$ is the infrared regulator mass. Propagators of the lowest components 
of the gauge superfield obtain a shift by the infrared regulator mass
which is enough to make the Feynman graphs safe in the infrared region of 
momentum space. It appears one can construct the classical action that satisfies 
a generalized ST identity which involves BRST counterparts of the gauge parameters. 
Then, a general solution to the generalized 
ST identity has been found at the classical level. As a consequence of 
that solution, the path integral that corresponds to this solution 
possesses the property of independence 
of v.e.v.s of gauge invariant quantities on the gauge parameters 
\cite{Piguet:1984im}. The independence of the physical quantities on the 
infrared scale was obtained 
by the same way. In addition, the new external superfield $u$ can be introduced 
so that a shift of its highest component is proportional to the infrared scale 
$\mu.$ That field also participates in the generalized ST identity.

One can analyse the generalized ST identity of Ref. \cite{Piguet:1984im} which 
is obtained after the modification of standard one (\ref{STrM}) by including 
additional external fields and gauge parameters. 
Appearance of the additional insertions of the external field $u$ or 
spurions $\mu {\t^2}{\bt^2}$ into supergraphs does not change the nonrenormalization 
theorems. This property has been used in Refs. 
\cite{Yamada:1994id,Jack:1997pa,Avdeev:1997vx}
to derive the relation 
between softly broken and rigid renormalization constants in ${\cal N} =1$ 
supersymmetric 
theory. Thus, it cannot bring any changes for our conclusions about the 
 $Lcc$ vertex from the point of 
our analysis since all divergent subgraphs remain divergent and all convergence 
properties of subgraphs remain unchanged. In that sense subgraphs are 
finite after renormalization by the dressing functions but all the vertex as a whole 
is also finite superficially due to property of Grassmannian 
integration which is not 
broken by the insertions of the external superfields 
\cite{Yamada:1994id,Jack:1997pa,Avdeev:1997vx}.

There is also another way to explain the independence of the physical quantities on the infrared scale $\mu.$ 
The point is that the the factors $(1 + \mu {\t^2}{\bt^2})$ coming from vertices 
will be canceled with factors $(1 + \mu {\t^2}{\bt^2})^{-1}$ coming from 
propagators. However, the propagators are IR-finite with the $\mu$ addition and thus 
the theory is regularized in the infrared. The same trick can be applied to demonstrate 
the independence of the correlators of dressed mean superfields in the 
Landau gauge of the infrared scale $\mu.$

As we have written at the end of the previous section, in the position space in component formalism the infrared divergences can be  analysed in the same way like ultraviolet divergences were analysed in Ref.\cite{Bogolyubov:1980nc} in momentum space 
by means of $R-$operation \cite{Cvetic:2006iu,Cvetic:2007ds}. The off-shell  IR divergence which has its  origin in the propagator of vector superfield is a typical problem for the superfield formalism  \cite{Piguet:1984mv,Piguet:1984im}.
 In the component formalism this problem does not exist, this statement may be checked by calculating index of divergence at infinity in  position space in complete analogy with 
 $R$-operation of the overlapping divergences in  momentum space   \cite{Cvetic:2006iu,Cvetic:2007ds}.

\section{Regularization of  UV divergence} 

In this paper we have significantly used the vanishing of the gauge beta function in ${\cal N}=4$ super-Yang-Mills theory. 
The vanishing of the beta function in first three orders of the perturbation theory has been established in Refs. 
\cite{Ferrara:1974pu,Jones:1977zr,Avdeev:1980bh}.  Originally, in the background field technique, it has been shown 
in Ref. \cite{Grisaru:1982zh} that in ${\cal N}=2$ supersymmetric Yang--Mills theory 
the beta function vanishes beyond one loop. The same result has been derived
in Ref. \cite{Howe:1983sr} by using the background field technique with unconstrained 
${\cal N}=2$ superfields. The arguments of \cite{Grisaru:1982zh,Howe:1983sr}  
are based on the fact that ${\cal N}=2$ supersymmetry prohibits any counterterms to the gauge coupling except 
for one loop contribution.  In Ref. 
\cite{West:1986wu}, the fact that ${\cal N}=2$ supersymmetric YM theory does 
not have contributions to the beta function beyond one loop has been 
argued based on currents of R-symmetry, which are in the same 
supermultiplet with the energy-momentum tensor. The  
proportionality of the trace anomaly of the energy momentum tensor to 
the beta function in general nonsupersymmetric Yang-Mills theories 
has been proved in Ref. \cite{Collins:1976yq}. Since R-symmetry does not have 
anomaly in ${\cal N}=4$ theory, the same is true for the anomaly of the 
conformal symmetry which is proportional to the beta function.        
At one loop level the beta function is zero with this field 
contents \cite{Ferrara:1974pu,Jones:1977zr}. Moreover, explicit calculation has 
been done at two loops in terms of ${\cal N}=1$ superfields \cite{Howe:1984xq},
and it has been shown that the beta function 
of ${\cal N}=2$ theory is zero at two loops.

The dimensional reduction was known to be inconsistent \cite{Siegel:1980qs}. We proposed solution to this
problem in component formalism in Ref. \cite{Cvetic:2006kk} for ${\cal N}=4$ super-Yang-Mills theory.
We can repeat similar arguments in case of superfield formalism. The only new feature here is the appearance 
of the infrared scale $\mu$ in subgraphs, as it has been explained above.  
The point is that the vertex $Lcc$ is always convergent superficially in superfield formalism. 
In ${\cal N} = 4$ super-Yang-Mills theory ultraviolet divergences in the subgraphs of 
the $Lcc$ vertex should cancel each other at the end. 
The insertion of the operator of the conformal anomaly into vacuum expectation values  of operators of 
gluonic fields at different points in spacetime is proportional to the beta function of the gauge coupling 
\cite{Collins:1976yq}. Due to the algebra of the four-dimensional supersymmetry the beta function should be zero 
\cite{Sohnius:1981sn}. Algebra of the supersymmetry operators in the Hilbert space created by dressed fields
can be considered as four-dimensional in Lorentz indices as well as in spinor indices since 
the limit $\e \rar 0$ is non-singular at one-loop order, two-loop order and higher orders 
as we have seen in the previous paragraphs. Thus, we can consider each correlator 
as pure four-dimensional, solving  order-by-order the problem of dimensional discrepancy of 
convolutions in Lorentz and spinor indices. The dependence on the infrared scale $\mu$ is canceled 
by itself from the contributions of vertices and propagators, as it has been shown above.

Another  regularization scheme for  ${\cal N}=1$  supersymmetric Yang--Mills theory exists
which is the higher derivatives regularization scheme \cite{West:1985jx,Krivoshchekov:1978xg}. 
According to Ref. \cite{West:1985jx}, ${\cal N}=2$ supersymmetry can be maintained by the regulator piece of the Lagrangian in HDR.
The scheme is discussed in detail in Ref. \cite{Faddeev:1980be} for the nonsupersymmetric case. A direct supersymmetric 
generalization of the regularization by higher derivatives  has been constructed in Ref. \cite{Krivoshchekov:1978xg}. 
This generalization has been considered in detail also in the paper \cite{West:1985jx}, in particular in the background 
field technique. As has been explicitly shown in Refs.  \cite{West:1985jx,Krivoshchekov:1978xg}, at one loop level HDR 
regularizes all the supergraphs in a gauge invariant manner and this repeats the 
corresponding construction  in the  nonsupersymmetric version of HDR \cite{Faddeev:1980be}.  However, when applied to
explicit examples, this approach is known to yield incorrect results in Landau gauge \cite{Martin:1994cg}. A number of 
proposals have been put forward to treat this problem \cite{Bakeyev:1996is,Asorey:1995tq}. As was shown in Ref.
\cite{Asorey:1995tq}, the contradiction, noticed in \cite{Martin:1994cg} is related
to the singular character of Landau gauge. In all other covariant gauges the method works and to include also the Landau 
gauge one has to add one more Pauli-Villars field to get the correct result \cite{Bakeyev:1996is,Asorey:1995tq}. 
Having used this regularization, new scheme has been proposed in  Refs. \cite{Slavnov:2003cx,Slavnov:2002kg,Slavnov:2001pu}. 
Calculations in the higher derivative regularization in terms of superfields can be found in 
Refs.\cite{Stepanyantz:2004sg,Soloshenko:2003nc,Soloshenko:2002np,Slavnov:2003cx,Slavnov:2002kg,Slavnov:2001pu}.   
To perform the calculation, it is proposed to break the gauge symmetry by using some HDR scheme with usual derivatives 
instead of covariant derivatives and then to restore the ST identity for the effective action by using some non-invariant 
counterterms. This problem has been solved in Refs. \cite{Slavnov:2003cx,Slavnov:2002kg,Slavnov:2001pu}.    
All the arguments, given above for the regularization by the dimensional reduction in favor of  finiteness of the
kernels of dressed mean superfields are valid also for the 
regularization by higher derivatives. The only difference is that to remove the regularization by higher derivatives 
 we take the limit $\Lambda \rar \infty$ instead of $\e \rar 0$ as it was for the case of dimensional reduction, where
$\Lambda$ is the regularization scale of HDR.

As we have already mentioned, the regularisation by higher derivatives  exists for ${\cal N}=2$ superfield formalism, ${\cal N}=1$ superfield formalism,  or for the components of the supermultiplets in the Wess-Zumino gauge.  In all these cases HDR may keep  ${\cal N}=4$ supersymmetry. This happens because the ${\cal N}=2$ SUSY regulators of \cite{West:1985jx} may be rewritten in terms of  ${\cal N}=1$  superfields 
and further in components in the Wess-Zumino gauge. Ward identities related with ${\cal N}=4$
supersymmetry (in particular with the multiplet of anomalies considered in the first paragraph of this Section) are valid in each of these formalisms under the higher derivative regularization. 
The vanishing of the gauge beta function may  be proved in each of these cases  
based on this connection between the trace anomaly and R-symmetry for ${\cal N}=4$ SYM. 
Despite that the renormalization constants do not coincide for the different members of the same multiplet in the Wess-Zumino  gauge \cite{Avdeev:1980bh}, all the Ward identities are valid in both the regularizations (DRED and HDR). After removing the regularizations the proper correlators of the dressed mean superfields  become 
exactly four-dimensional in terms of the dressed components or in terms of the dressed ${\cal N}=1$ superfields.

\section{Three-point Green function of dressed mean fields} \label{sec:three}

Conformal symmetry imposes strong restrictions on the Green functions \cite{Fradkin:1978pp}.
It is true for any conformal field theory in any dimensions. Fradkin and co-authors have constructed conformal invariant QED \cite{Fradkin:1978pp,Palchik:1982ta}. It was expected that QED in the infrared limit possesses the conformal invariance because the coupling becomes fixed in the IR limit due to the renormalization group equations. Renormalization procedure in the conformal QED has been analysed in \cite{Fradkin:1978pp,Palchik:1982ta}.  Conformal invariance would be expected in QCD in the UV limit too due to the renormalization group equations and asymptotic freedom. Thus, the conformal invariance has a sense in QCD in the UV limit and in QED in the IR limit. The three-point vector Green functions may be fixed in both the cases up to coefficients, the four-point Green functions are restricted too but not so strongly. There is a lot of freedom remained in the Green functions of higher rangs. In Ref. \cite{Palchik:1982ta}
in the framework of the conformal QED the connected Green functions of the gauge vector field  and spinor fields are constrained in position space  by the conformal symmetry.

We may apply these arguments of the Ref.  \cite{Palchik:1982ta} to the three-point connected Green function of the dressed mean vector superfields in order to find its form in position space up to coefficients. Because the effective action of the dressed mean fields does not depend on any scale, conformal transformations of the fields of the classical action
at the level of the effective action are converted to the corresponding conformal transformations of the dressed mean fields.  The generators of the conformal group will act on the dressed mean fields of the effective action. The trick is very similar to that how the ST identity imposed on the effective action has been obtained from the BRST symmetry of the classical action 
by means of the Legendre transformations \cite{Faddeev:1980be,Cvetic:2002dx} or even 
simpler example may be given, the invariance with respect to Lorentz symmetry may be obtained for the effective action via the Legendre transformation of the connected diagram generator. 
However, the conformal symmetry of the classical action restricts the proper vertices 
in an implicit way, but the constraints imposed by this symmetry on the connected Green functions 
are explicit.

In order to use the constraints from the conformal field theory the fields or superfields should be dynamical, otherwise we cannot transform the conformal symmetry of the classical action to the constraints on the vacuum expectation values of the fields (on the connected Green functions).  
By the same trick the Green functions of the softly broken and rigid supersymmetric theories were related in \cite{Kondrashuk:1999de}. This is the case of the dressed mean superfields from the 
Lagrangian of ${\cal N}=4$ SYM, they are dynamical. We may apply this trick to the connected correlators but not to a proper vertex. This is why we cannot fix the $Lcc$ vertex based on the conformal group transformations, the $L$ field is not dynamical, it is an external auxiliary source coupled to the composite double-ghost operator and not to the dynamical field 
from the integration measure of the path integral.

Thus, the conformal invariance imposes constraints on the connected  Green function of the dressed mean fields for all the orders in the gauge coupling. Nothing except for transformation laws of fields or operators with respect to the conformal group is necessary \cite{Fradkin:1978pp,Palchik:1982ta} to establish these restrictions. 
The three-point connected Green function is fixed up to coefficients. The coefficient is a series in terms of the  gauge coupling. Can it be checked explicitly by the direct calculations in position space? To calculate the connected Green function in position space at the tree  level it is necessary to integrate three propagators over coordinates of the common vertex. As we have mentioned in \cite{Allendes:2009bd}, for QED it has been done in Refs. \cite{Mitra:2008pw,Mitra:2008yr,Mitra:2009zm} in which  three classical propagators were integrated in the common vertex. The result is a structure predicted by the conformal group \cite{Mitra:2008pw,Mitra:2008yr,Mitra:2009zm}. In QCD the calculation at the tree level of the connected 
three propagators over the common vertex has been done 
in Ref. \cite{Cvetic:2007ds}. The result also satisfies  the form dictated by the conformal group. 
It is known that the structure of the integral of three propagators over the common vertex 
should contain the Davydychev integral $J(1,1,1)$ in position space \cite{Cvetic:2006iu}.  The conformal symmetry of the theory is the only reason that prohibits its appearance in the result for this  integral of three gluon propagators in the Landau gauge in position space
after summing all the contributions \cite{Cvetic:2007ds}.

In the previous paragraph of this section we wrote about the calculations in the component formalism at the tree level in QED and in ${\cal N}=4$ SYM for the connected Green function of three dressed gluons. The results are consistent with the constraints imposed by the conformal group.   
This conformal structure of the connected Green function should survive at the higher loops, however nobody has checked this by explicit calculations. In the literature there is no result (to our knowledge) about the calculation of the three point off-shell gluon function at two loops in QCD or   ${\cal N}=4$ SYM for a proper vertex or a connected Green correlator.  In Refs. \cite{Cvetic:2006iu,Allendes:2009bd} we wrote that the three-point connected Green functions of dressed gluons may be found explicitly at all the loops due to conformal symmetry in the Landau gauge in ${\cal N}=4$ SYM. We also made a conjecture in 
\cite{Cvetic:2006kk} that all the gauge dependence in other gauges should be in the dressing functions of the dressed mean fields in this theory. Calculation for the $Lcc$ vertex has been done explicitly at the two loop level in the component formalism \cite{Cvetic:2006iu,Cvetic:2007fp,Cvetic:2007ds}.

Now the question is why the structure of the $Lcc$ vertex looks so random  \cite{Cvetic:2006iu,Cvetic:2007fp,Cvetic:2007ds} while the structure of the connected three-gluon 
Green function of the dressed mean gluons is so simple and conformal? The $Lcc$ vertex is related by the ST identity to the proper (one particle irreducible)  three-gluon vertex diagram which has highly nontrivial structure and contains non-conformal $J(1,1,1)$ contributions \cite{Davydychev:1996pb} already at the one-loop level in momentum space for QCD. By construction of the dressed mean field correlators, this contribution should be summed with the one-loop self energy contribution
to the propagators to obtain the complete three-point connected Green function of the dressed gluons in the component formalism. As the result,  
these non-conformal $J(1,1,1)$ contributions should vanish in this total one-loop result for this Green function in momentum space for
${\cal N}=4$ SYM. This cancellation  may be possible because $J(1,1,1)$ is a highly non-trivial combination of the Appell functions and  its derivatives
that is resulted in a combination of Euler polylogarithms \cite{Davydychev:1992xr}. In momentum space this conformal  Green function consists of  a product of powers of the momenta of the propagator legs only.
Thus, the conformal structure is the only structure which survives in the connected three-point  Green function of the dressed gluons at the one loop level in position space and in momentum space. This structure does not contain non-conformal $J(1,1,1)$ contributions. Again, the only reason for cancellation  of these non-conformal $J(1,1,1)$ contributions at the one-loop level is the constraints imposed by the conformal symmetry in  position space on this connected correlator. It is consistent with the results of Refs.\cite{Kondrashuk:2008ec,Kondrashuk:2008xq,Kondrashuk:2009us,Allendes:2009bd} where it has been proven that 
three-point Green functions in the massless theories  are invariant with respect to Fourier transformation.

In the review \cite{Fradkin:1978pp} it is described how to use conformal invariance in order  to find restrictions on the correlation functions of gauge invariant operators in a conformal field theory for all the loops.  It appears to be a useful tool for work with operator product expansion in a conformal field theory. Later, this tool has been applied to AdS-CFT correspondence  \cite{Erdmenger:1996yc,Freedman:1998tz}. However, in this paper we consider the connected Green function of the dressed mean superfields in the Landau gauge.   
The one-loop structure for this $Lcc$ vertex in the component formalism is simple and coincides for any YM theory. It can be found in Refs.\cite{Cvetic:2006iu,Cvetic:2004kx,Cvetic:2006kk} . 
\begin{figure}[ht!] 
\centering\includegraphics[scale=0.6]{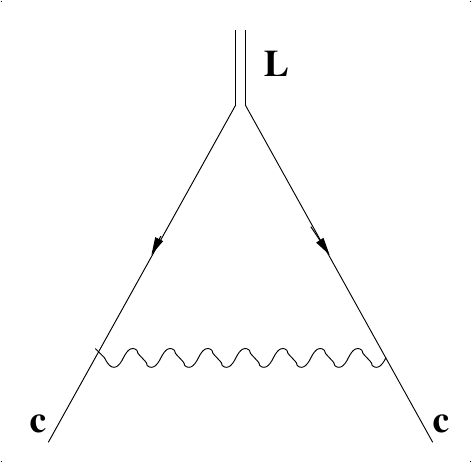}
\caption{\footnotesize  The only one-loop contribution to the $Lcc$ vertex}
\label{figure-1}
\end{figure}
The representation in terms of component fields is 
\begin{eqnarray*} 
\int~d^4x_1d^4x_2d^4x_3 \frac{i g^2 N}{2^{8}\pi^6} f^{abc} L^a(x_1)c^b(x_2)c^c(x_3) ~ V^{(1)} (x_1,x_2,x_3),
\end{eqnarray*}
where
\begin{eqnarray} \label{not}
V^{(1)}(x_1,x_2,x_3)  =  \no\\
\frac{-1}{[12]^2[23]^2} + \frac{2}{[12]^2[31]^2}  + \frac{-1}{[23]^2[31]^2}  + \frac{-1}{[12][23][31]^2} + \frac{2}{[12][23]^2[31]} +  \frac{-1}{[12]^2[23][31]}.   
\end{eqnarray}
We assume the concise notation of Ref.~\cite{Cvetic:2006iu},  where $[Ny]= (x_N - y)^2$ and analogously for $[Nz],$ and $[yz] = (y-z)^2$, 
that is, $N=1,2,3$ stands for $x_N=x_1,x_2,x_3$, respectively, which are the external points of the triangle diagram in Fig.\ref{figure-1} in position space. As we have mentioned in Ref. ~\cite{Cvetic:2006iu}, this representation (\ref{not})  does not match the representation for the connected scalar three-point function imposed by the conformal 
symmetry.

Moreover, the one-loop expression (\ref{not})  in position space is not written in a form 
convenient to solve the ST identity. We need a bit different representation to apply it in the ST identity. For this purpose, we first consider a scalar triangle diagram of Ref.\cite{Gonzalez:2018gch}
and note that any term in the one-loop contribution in $Lcc$ vertex (\ref{not}) 
may be considered as such a scalar triangle diagram depicted in Fig.\ref{figure-2} 
in position space with arbitrary indices $\alpha_1, \alpha_2,  \alpha_3$ on the propagators. 
\begin{figure}[ht!] 
\centering\includegraphics[scale=0.4]{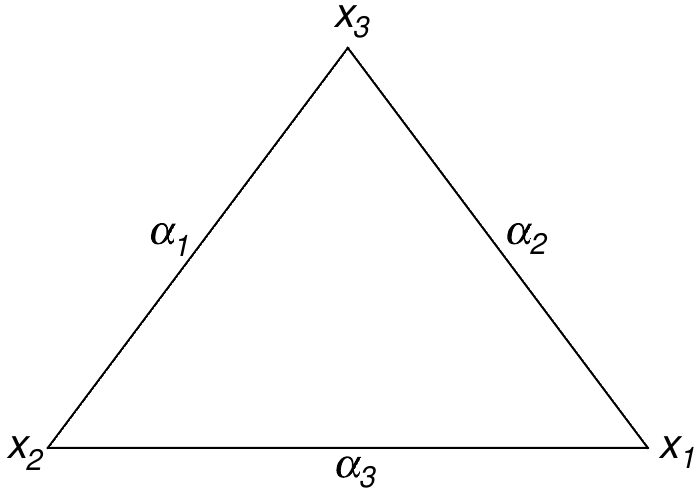}
\caption{\footnotesize  One-loop massless scalar triangle in position space. Integration in position space should be done over internal points only (See Ref. \cite{Allendes:2012mr}),
however this diagram does not have internal points at all. The point $x_1,$  $x_2,$  $x_3$ are external points.}
\label{figure-2}
\end{figure}
Let us first analyse the diagram in Fig.\ref{figure-2} in momentum space.  The $d$-dimensional momenta $p_1$, $p_2$, $p_3$ enter this diagram in  Fig.\ref{figure-1} and 
\begin{figure}[ht!] 
\centering\includegraphics[scale=0.5]{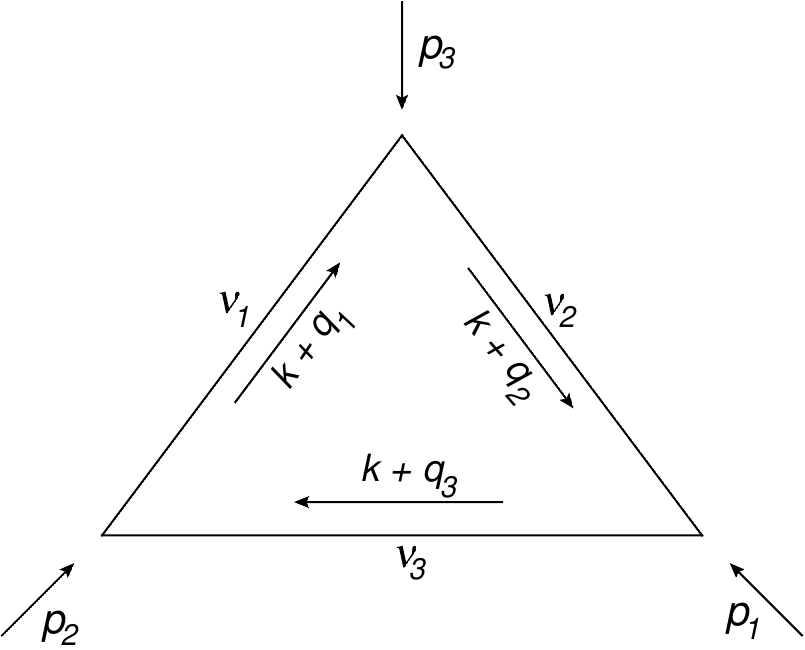}
\caption{\footnotesize  One-loop massless scalar triangle in momentum space}
\label{figure-3}
\end{figure}
are related by  momentum conservation
\begin{eqnarray*}
p_1 + p_2 + p_3 = 0.  
\end{eqnarray*} 
In momentum space  the diagram in Fig.\ref{figure-2} corresponds to the integral 
\begin{eqnarray} \label{J1}
J(\nu_1,\nu_2,\nu_3) = \int~Dk~\frac{1}{\left[(k + q_1)^2\right]^{\nu_1} \left[(k + q_2)^2 \right]^{\nu_2}
\left[(k + q_3)^2\right]^{\nu_3}}.
\end{eqnarray}
The running momentum $k$ in the triangle diagram in Fig.\ref{figure-2} is the integration variable in the integral (\ref{J1}).  The notation is chosen in such a way that the index of propagator $\nu_1$ stands on the line opposite to the vertex of triangle into which the momentum $p_1$ enters.  The notation $q_1,$ $q_2$ and $q_3$ are taken from Ref.\cite{Usyukina:1992jd}. As it may be seen from the diagram in Fig. \ref{figure-1} 
\begin{eqnarray*}
p_1 = q_3 - q_2, ~~~ 
p_2 = q_1 - q_3, ~~~
p_3 = q_2 - q_1. 
\end{eqnarray*}
The integral measure in momentum space is defined as in \cite{Cvetic:2006iu} by 
\begin{eqnarray*}
Dk \equiv \pi^{-\frac{d}{2}}d^d k.    
\end{eqnarray*} 
Such a definition of the integration measure in momentum space helps to avoid powers of $\pi$ in formulas for the momentum integrals. The Mellin-Barnes representation of the integral $J(\nu_1,\nu_2,\nu_3)$ may be obtained via Feynman parameters \cite{Usyukina:1992jd} and has the form 
\begin{eqnarray*}
J(\nu_1,\nu_2,\nu_3) = \frac{1}{\Pi_{i} \G(\nu_i) \G(d-\Sigma_i \nu_i)} \frac{1}{{(p^2_3)}^{ \Sigma \nu_i -d/2}}
\oint_C dz_2~dz_3 \le\frac{p^2_1}{p^2_3}\ri^{z_2}  \le\frac{p^2_2}{p^2_3}\ri^{z_3} 
\left\{ \G \le -z_2 \ri \G \le -z_3 \ri \right.\no\\
\left. \G \le -z_2 -\nu_2-\nu_3 + d/2 \ri \G \le -z_3-\nu_1-\nu_3 + d/2 \ri 
\G \le z_2 + z_3  + \nu_3 \ri  \G \le  \Sigma \nu_i - d/2 + z_3 + z_2 \ri \right\}  \\
\equiv \frac{1}{{(p^2_3)}^{ \Sigma \nu_i -d/2}}\oint_C dz_2~dz_3 x^{z_2}  y^{z_3} 
D^{(z_2,z_3)} [\nu_1,\nu_2,\nu_3]. \no
\end{eqnarray*}
We have used here the definition of the Mellin-Barnes transform $D^{(z_2,z_3)}[\nu_1,\nu_2,\nu_3]$ from our Ref.\cite{Allendes:2012mr}, 
\begin{eqnarray*}
D^{(z_2,z_3)}[\nu_1,\nu_2,\nu_3] = \frac{ \G \le -z_2 \ri \G \le -z_3 \ri \G \le -z_2 -\nu_2-\nu_3 + d/2 \ri 
\G \le -z_3-\nu_1-\nu_3 + d/2 \ri }
{\Pi_{i} \G(\nu_i) } \no\\
\times 
\frac{ \G \le z_2 + z_3  + \nu_3 \ri  \G \le  \Sigma \nu_i - d/2 + z_3 + z_2 \ri }
{\G(d-\Sigma_i \nu_i)}.
\end{eqnarray*}
With all these definitions made, we consider the triangle scalar diagram in Fig.  \ref{figure-2}
in position space, repeating in part the integral transformations used in Ref. \cite{Gonzalez:2018gch}. 

\begin{eqnarray*}
\frac{1}{[12]^{\a_3}[23]^{\a_1}[31]^{\a_2}} = \frac{  \pi^{-3d/2} 4^{-\Sigma_i \a_i}  \G(d/2-\a_1)\G(d/2-\a_2) \G(d/2-\a_3)}{\G(\a_1)\G(\a_2)\G(\a_3)} \times\\
\int¬dq_1dq_2dq_3 \frac{e^{iq_3(x_1-x_2)} e^{iq_1(x_2-x_3)} e^{iq_2(x_3-x_1)}}{(q_3^2)^{d/2-\a_3}(q_1^2)^{d/2-\a_1} (q_2^2)^{d/2-\a_2}}   \\
= \frac{\pi^{-3d/2} 4^{-\Sigma_i \a_i}\G(d/2-\a_1)\G(d/2-\a_2) \G(d/2-\a_3)}{\G(\a_1)\G(\a_2)\G(\a_3)} 
\int¬dq_1dq_2dq_3 \frac{e^{ix_1(q_3-q_2)} e^{ix_2(q_1-q_3)} e^{ix_3(q_2-q_1)}}{(q_3^2)^{d/2-\a_3}(q_1^2)^{d/2-\a_1} (q_2^2)^{d/2-\a_2}} \\ 
= \frac{\pi^{-3d/2} 4^{-\Sigma_i \a_i}\G(d/2-\a_1)\G(d/2-\a_2) \G(d/2-\a_3)}{\G(\a_1)\G(\a_2)\G(\a_3)} \times\\  
\int¬dp_1dp_2dp_3dq_3\delta(p_1+p_2+p_3)    \frac{e^{ix_1p_1} e^{ix_2p_2} e^{ix_3p_3}}{[q_3^2]^{d/2-\a_3}[(p_2+q_3)^2]^{d/2-\a_1} [(p_1-q_3)^2]^{d/2-\a_2}} \\
= \frac{\pi^{-3d/2} 4^{-\Sigma_i \a_i}\G(d/2-\a_1)\G(d/2-\a_2) \G(d/2-\a_3)}{(2\pi)^d\G(\a_1)\G(\a_2)\G(\a_3)} \times\\  
\int¬dp_1dp_2dp_3dq_3dx_5    \frac{e^{i(x_1-x_5)p_1} e^{i(x_2-x_5)p_2} e^{i(x_3-x_5)p_3}}{[q_3^2]^{d/2-\a_3}[(p_2+q_3)^2]^{d/2-\a_1} [(p_1-q_3)^2]^{d/2-\a_2}} \\
= \frac{\pi^{-d} 4^{-\Sigma_i \a_i}\G(d/2-\a_1)\G(d/2-\a_2) \G(d/2-\a_3)}{(2\pi)^d\G(\a_1)\G(\a_2)\G(\a_3)} \times\\  
\int¬dp_1dp_2dp_3dx_5  e^{i(x_1-x_5)p_1} e^{i(x_2-x_5)p_2} e^{i(x_3-x_5)p_3} J(d/2-\a_1,d/2-\a_2,d/2-\a_3) \\
= \frac{\pi^{-d} 4^{-\Sigma_i \a_i}\G(d/2-\a_1)\G(d/2-\a_2) \G(d/2-\a_3)}{(2\pi)^d\G(\a_1)\G(\a_2)\G(\a_3)} \times\\
\int¬dp_1dp_2dp_3dx_5 \oint_C~dz_2dz_3  \frac{e^{i(x_1-x_5)p_1} e^{i(x_2-x_5)p_2} e^{i(x_3-x_5)p_3}} {(p_3^2)^{d-\Sigma_i \a_i +z_2+z_3}(p_1^2)^{-z_2} (p_2^2)^{-z_3} } 
D^{(z_2,z_3)}[d/2-\a_1,d/2-\a_2,d/2-\a_3] \\
= \frac{\pi^{d/2} 4^{d/2}\G(d/2-\a_1)\G(d/2-\a_2) \G(d/2-\a_3)}{(2\pi)^d\G(\a_1)\G(\a_2)\G(\a_3)} \times\\  
\int¬dx_5 \oint_C~dz_2dz_3  \frac{D^{(z_2,z_3)}[d/2-\a_1,d/2-\a_2,d/2-\a_3]} {[35]^{\Sigma_i \a_i -z_2-z_3-d/2}[15]^{d/2+z_2} [25]^{d/2+z_3} } \times\\
\frac{\G(\Sigma_i \a_i -z_2-z_3-d/2 )\G(d/2 + z_2) \G(d/2+ z_3)}{\G(d-\Sigma_i \a_i +z_2 + z_3 )\G(-z_2 )\G(-z_3)}. 
\end{eqnarray*}
Any term in (\ref{not}) may be represented in this form, and for the all one-loop $Lcc$ vertex in the component formalism we may write  
\begin{eqnarray} \label{param}
\int~d^4x_1d^4x_2d^4x_3 \frac{i g^2 N}{2^{8}\pi^6} f^{abc} L^a(x_1)c^b(x_2)c^c(x_3) ~ V^{(1)} (x_1,x_2,x_3) \no\\
= \int~d^4x_1d^4x_2d^4x_3 \frac{i g^2 N}{2^{8}\pi^6} f^{abc} L^a(x_1)c^b(x_2)c^c(x_3)
\int¬dx_5 \oint_C~dz_2dz_3  \frac{{\cal M}(z_2,z_3)} {[15]^{\Delta-z_2-z_3} [25]^{z_2} [35]^{z_3} }. 
\end{eqnarray}
Here $\Delta$  is the dimension of this $Lcc$ vertex. In the case that we consider ( ${\cal N}=4$ SYM in $d=4$)  $\Delta=4.$ The Eq. (\ref{param}) is the parametrization for the $Lcc$ vertex we need to solve the ST identity by the method developed in Refs. \cite{Cvetic:2002dx,Cvetic:2002in}. This parametrization is valid for any number of loops. This auxiliary vertex is related to the three-gluon proper vertex via ST identity in the way we described in Section \ref{sec:from}. Thus, in the component formalism the conformal structure of the connected Green function of three dressed gluons transforms  implicitly to the structure of the $Lcc$ vertex because the integral relation between the connected Green function of dressed gluons and the corresponding proper function of dressed gluons is highly nontrivial in all the loops. In turn,  the Bethe-Salpeter  equation puts strong restrictions on  
the Mellin-Barnes image ${\cal M}(z_2,z_3)$ of the $Lcc$ vertex in the parametrization (\ref{param})  
in an explicit way \cite{Allendes:2009bd,Allendes:2012mr}.

The $Lcc$ is finite in the limit of removing the regularizations as we have shown in Refs. 
\cite{Cvetic:2004kx,Cvetic:2006kk} and checked explicitly in Refs.\cite{Cvetic:2006iu,Cvetic:2007fp,Cvetic:2007ds} in the component formalism. We have written in the Introduction that scale independence is a consequence of finiteness. Now, we do not have any scale because the amplitudes
are on-shell values of the connected Green functions of dressed fields and therefore IR divergences after putting the moments  on-shell may appear. In order to work with these connected correlators on shell we need a regularization again. This may be completely new regularization, for example, we may shift the 
complex variables of integration in the Mellin planes by some value $\epsilon$ and obtain 
\begin{eqnarray*} 
\int~d^4x_1d^4x_2d^4x_3 \frac{i g^2 N}{2^{8}\pi^6} f^{abc} L^a(x_1)c^b(x_2)c^c(x_3)
\int¬dx_5 \oint_C~dz_2dz_3  \frac{{\cal M}(z_2 + \epsilon,z_3+\epsilon)} {[15]^{\Delta-z_2-z_3 -2\epsilon} [25]^{z_2  +\epsilon} [35]^{z_3  +\epsilon} } 
\end{eqnarray*}
instead of (\ref{param}). Such a regularization will regularize the IR divergences on-shell. 
 We do not need any regularization of such a kind at the tree level. In terms of this vertex all other vertices may be fixed  \cite{Cvetic:2002dx,Cvetic:2002in,Allendes:2009bd,Allendes:2012mr,Borja:2016sap}.

\section{Chiral superfields}

So far the pure gauge sector has been considered. Let us look now at the matter 
two point functions. Schematically, one can write the two point vertex as
\begin{eqnarray*}
\int d^8zd^8z'\bF(z)G_\F(z-z')\F(z').
\end{eqnarray*} 
The function $G_\F(z-z')$ can be divided into two equal parts $\tilde{G}_\F,$
\begin{eqnarray*}
\int d^8z'\tilde{G}_\F(z_1-z')\tilde{G}_\F(z'-z_2) = G_\F(z_1-z_2).
\end{eqnarray*} 
This is a product in momentum space. 
Now we define the
new fields $\tilde{\F},$ 
\begin{eqnarray*}
\int d^8z'\tilde{G}_\F(z-z')\F(z') \equiv  \tilde{\F}(z),
\end{eqnarray*} 
and represent the effective action in terms of these fields. In particular,
the divergent part of the function  $\tilde{G}_\F$ can be absorbed into the 
redefinition of the Yukawa couplings and masses. However, in  ${\cal N}=4$
supersymmetric theory masses are absent and the Yukawa coupling (which coincides with 
the gauge coupling) is not renormalized due to the structure of the Yukawa terms
in the classical action. Thus, in terms of the dressed effective superfields 
$\tilde{\F}$ and $\Vt$ the effective action does not have any dependence on the 
UV and IR scales for ${\cal N}=4$ supersymmetric theory.

Because the  superficial divergence in the Yukawa coupling  
for ${\cal N}=1$ supersymmetry is absent due to Grassmannian integration in the superspace \cite{West:1986wu},
all the renormalization from the self-energy of the chiral superfields should be absorbed by the 
renormalization of the 
Yukawa coupling. But in this special case of ${\cal N}=4$
SYM the gauge coupling and Yukawa coupling is the same and gauge coupling does not get any renormalization in this theory due to the vanishing of the gauge beta function, this means that the chiral superfields in this model should not be renormalized in order to absorb infinities, that is, the self energy of the chiral superfields may be finite only.  In contrary, in the component formalism self-energies of components of the chiral multiplet are divergent \cite{Minahan:2002ve}. 

Absence of the renormalization of the self-energy for the chiral superfields is not the only example when the bubble diagram does not contribute. Another example would be the massless 
bubble diagrams with external on-shell momenta of Ref.\cite{Arkani-Hamed:2008owk} which are divergent both in UV and IR, but they vanish due to cancellation between UV and IR poles in $d = 4-2\epsilon$. It implies the IR and UV divergences are related to each other and can be determined from one another \cite{Arkani-Hamed:2008owk}.

\section{Conclusion}

In conclusion, all the correlator $Lcc$ with all the possible contributions included turns out to be totally finite in 
${\cal N}=4$ super Yang-Mills theory in the Landau gauge and this property can be used to find it exactly in all the orders 
of the perturbation theory. The vertex $Lcc$ in spite of being scale independent cannot be found by conformal symmetry 
since the external auxiliary superfield $L$ does not propagate (it is not in the measure of the path integral). 
Three point connected Green functions of supermultiplets containing  physical fields (like vector 
supermultiplet or matter supermultiplet) could be fixed  by conformal symmetry up to some coefficient depending on the 
gauge coupling and number of colours.

In QCD the matter fields are not in adjoint representation, the level of the symmetry is much lower 
and the beta function does not vanish. The rest of  singularity in the vector propagator  
can be absorbed into the gauge coupling to organize the bare coupling. The bare coupling together with the logarithm of ratio
of the distance to the scale, leads to the running (effective) coupling. It means that in $d$
dimensions the massless nonsupersymmetric gauge theory is a conformal gauge theory in
terms of the running effective coupling (formed from the bare coupling) and dressed mean
fields \cite{Cvetic:2007fp}. Another way to break supersymmetry and obtain the results for QCD 
from supersymmetric theories is via diffeomorphisms in the superspace \cite{Kondrashuk:1999de}.

\subsection*{Acknowledgments}

The work of I.K. was supported by Ministry of Education (Chile) under grant 
Mecesup FSM9901 and by DGIP UTFSM, by Fondecyt (Chile) Grants Nos. 1040368, 1050512 and 1121030, by DIUBB (Chile) Grant Nos.  125009,  GI 153209/C  and GI 152606/VC.
The work of I.S. is supported by grants ANID-Chile FONDECYT 1230391 and  ANID PIA/APOYO AFB220004.
We are grateful to Gorazd Cveti\v{c} for many useful discussions. 
These results were presented in the talk of I.K. at XIV symposium of Chilean society of physics (Sochifi), Antofagasta, Chile, November 16-19, 2004. He is grateful to Marcelo Loewe, Miguel Calvo and Ivan Gonzalez  for their interest and questions during this talk.  We thank all four 
referees of the present paper for their comments ans suggestions which  improved it significantly.


\begin{thebibliography}{99}



\bibitem{Slavnov:1972fg}
A.~A.~Slavnov,
``Ward Identities In Gauge Theories,''
Theor.\ Math.\ Phys.\  {\bf 10} (1972) 99
[Teor.\ Mat.\ Fiz.\  {\bf 10} (1972) 153].


\bibitem{Taylor:1971ff}
J.~C.~Taylor,
``Ward Identities And Charge Renormalization Of The Yang-Mills Field,''
Nucl.\ Phys.\ B {\bf 33} (1971) 436.



\bibitem{Slavnov:1974dg}
A.~A.~Slavnov,
``Renormalization Of Supersymmetric Gauge Theories. 2. Nonabelian Case,''
Nucl.\ Phys.\ B {\bf 97} (1975) 155.


\bibitem{Faddeev:1980be}
L.~D.~Faddeev and A.~A.~Slavnov,
{\it Gauge Fields. Introduction To Quantum Theory,} Frontiers in Physics Series, Vol. {\bf 50}, (Benjamin/Cummings, Reading, Massachusetts, 1980), p.1;
 Frontiers in Physics Series, Vol. {\bf 83}, (Addison-Wesley, Redwood City, California, 1990), p.1; 
{\it Introduction to quantum theory of gauge fields,} (Nauka, Moscow,  1988).







\bibitem{Lee:1973hb}
B.~W.~Lee,
``Transformation Properties Of Proper Vertices In Gauge Theories,''
Phys.\ Lett.\ B {\bf 46} (1973) 214.



\bibitem{Zinn-Justin:1974mc}
J.~Zinn-Justin, ``Renormalization Of Gauge Theories,'' Trends in Elementary Particle Theory, Springer Book Series: Lecture Notes in Physics, Vol. {\bf 37}, edited by
H. Rollnik and K. Dietz, (Springer, Berlin/Heidelberg, Germany, 1975), pp. 1-39. 






\bibitem{Becchi:1974md}
C.~Becchi, A.~Rouet and R.~Stora,
``Renormalization Of The Abelian Higgs-Kibble Model,''
Commun.\ Math.\ Phys.\  {\bf 42} (1975) 127.




\bibitem{Tyutin:1975qk}
I.~V.~Tyutin,
``Gauge Invariance In Field Theory And Statistical Physics In Operator Formalism,''
LEBEDEV-75-39 (in Russian), 1975. 





\bibitem{Piguet:1996ys}
O.~Piguet, 
``Supersymmetry, supercurrent, and scale invariance,''
arXiv:hep-th/9611003.



\bibitem{Krivoshchekov:1978xg}
V.~K.~Krivoshchekov,
``Invariant Regularizations For Supersymmetric Gauge Theories,''
Teor.\ Mat.\ Fiz.\  {\bf 36} (1978) 291.



\bibitem{West:1985jx}
P.~C.~West,
``Higher Derivative Regulation Of Supersymmetric Theories,''
Nucl.\ Phys.\ B {\bf 268} (1986) 113.




\bibitem{Siegel:1979wq}
W.~Siegel,
``Supersymmetric Dimensional Regularization Via Dimensional Reduction,''
Phys.\ Lett.\ B {\bf 84}, 193 (1979).




\bibitem{Capper:1979ns}
D.~M.~Capper, D.~R.~T.~Jones and P.~van Nieuwenhuizen,
``Regularization By Dimensional Reduction Of Supersymmetric And
Nonsupersymmetric Gauge Theories,''
Nucl.\ Phys.\ B {\bf 167} (1980) 479.



\bibitem{Cvetic:2004kx}
  G.~Cvetic, I.~Kondrashuk and I.~Schmidt,
   ``Effective action of dressed mean fields for N = 4 super-Yang-Mills theory,''
  Mod.\ Phys.\ Lett.\ A {\bf 21} (2006) 1127
  [arXiv:hep-th/0407251].





\bibitem{Cvetic:2006kk}
G.~Cveti\v{c}, I.~Kondrashuk and I.~Schmidt,
``On the effective action of dressed mean fields for N = 4 super-Yang-Mills 
theory,'' in {\it Symmetry, Integrability and Geometry: Methods and 
Applications,} SIGMA (2006) 002, arXiv:math-ph/0601002.



\bibitem{Cvetic:2006iu}
G.~Cvetic, I.~Kondrashuk, A.~Kotikov and I.~Schmidt,
``Towards the two-loop Lcc vertex in Landau gauge,''
Int. J. Mod. Phys. A \textbf{22} (2007), 1905-1934
[arXiv:hep-th/0604112 [hep-th]].









\bibitem{Sohnius:1981sn}
M.~F.~Sohnius and P.~C.~West,
``Conformal Invariance In N=4 Supersymmetric Yang-Mills Theory,''
Phys.\ Lett.\ B {\bf 100} (1981) 245.





\bibitem{Cvetic:2002dx}
I.~Kondrashuk, G.~Cveti\v{c},    and I.~Schmidt,
``Approach to solve Slavnov-Taylor identities in nonsupersymmetric non-Abelian 
gauge theories,''
Phys.\ Rev.\ D {\bf 67} (2003) 065006
[arXiv:hep-ph/0203014].


\bibitem{Cvetic:2002in}
G.~Cveti\v{c}, I.~Kondrashuk and I.~Schmidt,
``QCD effective action with dressing functions: Consistency checks in
perturbative regime,''
Phys.\ Rev.\ D {\bf 67} (2003) 065007
[arXiv:hep-ph/0210185].



\bibitem{Kondrashuk:2000br}
I.~Kondrashuk,
``The solution to Slavnov-Taylor identities in D4 N = 1 SYM,''
JHEP {\bf 0011}, 034 (2000)
[arXiv:hep-th/0007136].





\bibitem{Kondrashuk:2003tw}
  I.~Kondrashuk,
  ``An approach to solve Slavnov-Taylor identity in D4 N = 1 supergravity,''
  Mod.\ Phys.\ Lett.\ A {\bf 19} (2004) 1291
  [arXiv:gr-qc/0309075].




\bibitem{Kang:2004cs}
  K.~Kang and I.~Kondrashuk, ``Semiclassical scattering amplitudes of dressed gravitons,''
  arXiv:hep-ph/0408168.








\bibitem{Kondrashuk:2001yd} I.~Kondrashuk, ``The solution to Slavnov-Taylor identities in a general four dimensional supersymmetric 
theory,'' arXiv:hep-th/0110045.

  
  
  
  

\bibitem{Kinoshita:1962ur}
T.~Kinoshita,
``Mass singularities of Feynman amplitudes,''
J. Math. Phys. \textbf{3} (1962), 650-677



\bibitem{Alvarez:2016juq}
G.~Alvarez, G.~Cvetic, B.~A.~Kniehl, I.~Kondrashuk and I.~Parra-Ferrada,
``Analytical Solution to DGLAP Integro-Differential Equation in a Simple Toy-Model with a Fixed Gauge Coupling,''
Quantum Rep. \textbf{5} (2023) no.1, 198-223
[arXiv:1611.08787 [hep-ph]].



\bibitem{Bianchi:2013sta}
L.~Bianchi, V.~Forini and A.~V.~Kotikov,
``On DIS Wilson coefficients in N=4 super Yang-Mills theory,''
Phys. Lett. B \textbf{725} (2013), 394-401
[arXiv:1304.7252 [hep-th]].


\bibitem{Alvarez:2019eaa}
G.~Alvarez and I.~Kondrashuk,
``Analytical solution to DGLAP integro-differential equation via complex maps in domains of contour integrals,''
J. Phys. Comm. \textbf{4} (2020) no.7, 075004
[arXiv:1912.02303 [hep-th]].


\bibitem{Kondrashuk:2019cwi}
I.~Kondrashuk,
``Algorithm to find an all-order in the running coupling solution to an equation of the DGLAP type,''
Phys. Part. Nucl. Lett. \textbf{18} (2021) no.2, 141-147
[arXiv:1906.07924 [hep-ph]].



  
\bibitem{Minahan:2002ve}
J.~A.~Minahan and K.~Zarembo,
``The Bethe ansatz for N=4 superYang-Mills,''
JHEP \textbf{03} (2003), 013
[arXiv:hep-th/0212208 [hep-th]].
  

  
  
  
  
  
  
  
  
  
  
  
  
  


\bibitem{Kondrashuk:2000qb}
I.~Kondrashuk,
``Renormalizations in softly broken N = 1 theories: Slavnov-Taylor
identities,''
J.\ Phys.\ A {\bf 33}, 6399 (2000)
[arXiv:hep-th/0002096].



\bibitem{Gates:1983nr}
S.~J.~Gates, M.~T.~Grisaru, M.~Rocek and W.~Siegel,
``Superspace, Or One Thousand And One Lessons In Supersymmetry,''
Front.\ Phys.\  {\bf 58}, 1 (1983)
[arXiv:hep-th/0108200].







\bibitem{Piguet:1984mv}
  O.~Piguet and K.~Sibold,
  ``Gauge Independence In N=1 Supersymmetric Yang-Mills Theories,''
  Nucl.\ Phys.\ B {\bf 248} (1984) 301.



\bibitem{Piguet:1984im}
  O.~Piguet and K.~Sibold,
  ``The Off-Shell Infrared Problem In N=1 Supersymmetric Yang-Mills Theories,''
  Nucl.\ Phys.\ B {\bf 248} (1984) 336.


  
\bibitem{Grisaru:1979wc}
M.~T.~Grisaru, W.~Siegel and M.~Rocek,
``Improved Methods For Supergraphs,''
Nucl.\ Phys.\ B {\bf 159} (1979) 429.
  


\bibitem{West:1986wu}
P.~C.~West,
``Introduction To Supersymmetry And Supergravity,''
World Scientific (1986).

  
  
  
\bibitem{Avdeev:1992jt}
L.~V.~Avdeev, D.~I.~Kazakov and I.~N.~Kondrashuk,
``Renormalizations in supersymmetric and nonsupersymmetric nonAbelian Chern-Simons field theories with matter,''
Nucl. Phys. B \textbf{391} (1993), 333-357
  
  
  
  

\bibitem{Bogolyubov:1980nc}
N.~N.~Bogolyubov and D.~V.~Shirkov,
``Introduction To The Theory Of Quantized Fields,''
Intersci.\ Monogr.\ Phys.\ Astron.\  {\bf 3}, 1 (1959).

  
  






\bibitem{Yamada:1994id}
  Y.~Yamada,
  ``Two loop renormalization group equations for soft SUSY breaking scalar
  interactions: Supergraph method,''
  Phys.\ Rev.\ D {\bf 50} (1994) 3537
  [arXiv:hep-ph/9401241].



\bibitem{Jack:1997pa}
  I.~Jack and D.~R.~T.~Jones,
  ``The gaugino beta-function,''
  Phys.\ Lett.\ B {\bf 415} (1997) 383
  [arXiv:hep-ph/9709364].



\bibitem{Avdeev:1997vx}
  L.~V.~Avdeev, D.~I.~Kazakov and I.~N.~Kondrashuk,
  ``Renormalizations in softly broken SUSY gauge theories,''
  Nucl.\ Phys.\ B {\bf 510} (1998) 289
  [arXiv:hep-ph/9709397].


\bibitem{Cvetic:2007ds}
G.~Cvetic and I.~Kondrashuk,
``Gluon self-interaction in the position space in Landau gauge,''
Int. J. Mod. Phys. A \textbf{23} (2008), 4145-4204
[arXiv:0710.5762 [hep-th]].











\bibitem{Ferrara:1974pu}
S.~Ferrara and B.~Zumino,
``Supergauge Invariant Yang-Mills Theories,''
Nucl.\ Phys.\ B {\bf 79} (1974) 413.



\bibitem{Jones:1977zr}
D.~R.~T.~Jones,
``Charge Renormalization In A Supersymmetric Yang-Mills Theory,''
Phys.\ Lett.\ B {\bf 72} (1977) 199.




\bibitem{Avdeev:1980bh}
L.~V.~Avdeev, O.~V.~Tarasov and A.~A.~Vladimirov,
``Vanishing Of The Three Loop Charge Renormalization Function In A
Supersymmetric Gauge Theory,''
Phys.\ Lett.\ B {\bf 96} (1980) 94.



















\bibitem{Grisaru:1982zh}
M.~T.~Grisaru and W.~Siegel,
``Supergraphity. 2. Manifestly Covariant Rules And Higher Loop Finiteness,''
Nucl.\ Phys.\ B {\bf 201}, 292 (1982)
[Erratum-ibid.\ B {\bf 206}, 496 (1982)].





\bibitem{Howe:1983sr}
P.~S.~Howe, K.~S.~Stelle and P.~K.~Townsend,
``Miraculous Ultraviolet Cancellations In Supersymmetry Made Manifest,''
Nucl.\ Phys.\ B {\bf 236}, 125 (1984).












\bibitem{Collins:1976yq}
J.~C.~Collins, A.~Duncan and S.~D.~Joglekar,
``Trace And Dilatation Anomalies In Gauge Theories,''
Phys.\ Rev.\ D {\bf 16}, 438 (1977).




\bibitem{Howe:1984xq}
P.~S.~Howe and P.~C.~West,
``The Two Loop Beta Function In Models With Extended Rigid Supersymmetry,''
Nucl.\ Phys.\ B {\bf 242}, 364 (1984).





\bibitem{Siegel:1980qs}
  W.~Siegel, ``Inconsistency Of Supersymmetric Dimensional Regularization,''
  Phys.\ Lett.\ B {\bf 94} (1980) 37.






\bibitem{Martin:1994cg}
C.~P.~Martin and F.~Ruiz Ruiz,
``Higher covariant derivative Pauli-Villars regularization does not
 lead to a consistent QCD,''
Nucl.\ Phys.\ B  {\bf  436} (1995) 545
[arXiv:hep-th/9410223].


\bibitem{Bakeyev:1996is}
  T.~D.~Bakeyev and A.~A.~Slavnov,
  ``Higher covariant derivative regularization revisited,''
Mod.\ Phys.\ Lett.\ A {\bf 11} (1996) 1539
  [arXiv:hep-th/9601092].


\bibitem{Asorey:1995tq}
M.~Asorey and F.~Falceto,
``On the consistency of the regularization of gauge theories by high covariant
derivatives,''
Phys.\ Rev.\ D {\bf 54} (1996) 5290
[arXiv:hep-th/9502025].




\bibitem{Slavnov:2003cx}
A.~A.~Slavnov and K.~V.~Stepanyantz,
``Universal invariant renormalization of supersymmetric Yang-Mills  theory,''
Theor.\ Math.\ Phys.\  {\bf 139} (2004) 599
[Teor.\ Mat.\ Fiz.\  {\bf 139} (2004) 179]
[arXiv:hep-th/0305128].




\bibitem{Slavnov:2002kg}
A.~A.~Slavnov and K.~V.~Stepanyantz,
``Universal invariant renormalization for supersymmetric theories,''
Theor.\ Math.\ Phys.\  {\bf 135} (2003) 673
[Teor.\ Mat.\ Fiz.\  {\bf 135} (2003) 265]
[arXiv:hep-th/0208006].




\bibitem{Slavnov:2001pu}
A.~A.~Slavnov,
``Universal gauge invariant renormalization,''
Phys.\ Lett.\ B {\bf 518} (2001) 195.




\bibitem{Stepanyantz:2004sg}
K.~V.~Stepanyantz,
``Investigation of the anomaly puzzle in N = 1 supersymmetric electrodynamics,''
Theor. Math. Phys. \textbf{142} (2005), 29-47
[arXiv:hep-th/0407201 [hep-th]].


\bibitem{Soloshenko:2003nc}
A.~A.~Soloshenko and K.~V.~Stepanyantz,
``Three loop beta function for N=1 supersymmetric electrodynamics, regularized by higher derivatives,''
Theor. Math. Phys. \textbf{140} (2004), 1264-1282
[arXiv:hep-th/0304083 [hep-th]].





\bibitem{Soloshenko:2002np}
A.~Soloshenko and K.~Stepanyantz,
``Two-loop renormalization of N = 1 supersymmetric electrodynamics,
regularized by higher derivatives,''
arXiv:hep-th/0203118.




\bibitem{Fradkin:1978pp}
E.~S.~Fradkin and M.~Y.~Palchik,
``Recent Developments in Conformal Invariant Quantum Field Theory,''
Phys. Rept. \textbf{44} (1978), 249-349



\bibitem{Palchik:1982ta}
M.~Y.~Palchik,
``A New Approach to the Conformal Invariance Problem in Quantum Electrodynamics,''
J. Phys. A \textbf{16} (1983), 1523


\bibitem{Kondrashuk:1999de}
I.~Kondrashuk,
``On the relation between Green functions of the SUSY theory with and without soft terms,''
Phys. Lett. B \textbf{470} (1999), 129-133
[arXiv:hep-th/9903167 [hep-th]].


\bibitem{Allendes:2009bd}
P.~Allendes, N.~Guerrero, I.~Kondrashuk and E.~A.~Notte Cuello,
``New four-dimensional integrals by Mellin-Barnes transform,''
J. Math. Phys. \textbf{51} (2010), 052304
[arXiv:0910.4805 [hep-th]].


\bibitem{Mitra:2008yr}
I.~Mitra,
``On conformal invariant integrals involving spin one-half and spin-one particles,''
J. Phys. A \textbf{41} (2008), 315401
[arXiv:0803.2630 [hep-th]].


\bibitem{Mitra:2008pw}
I.~Mitra,
``Three-point Green function of massless QED in position space to lowest order,''
J. Phys. A \textbf{42} (2009), 035404
[arXiv:0808.2448 [hep-th]].


\bibitem{Mitra:2009zm}
I.~Mitra,
``External leg amputation in conformal invariant three-point function,''
Eur. Phys. J. C \textbf{71} (2011), 1621
[arXiv:0907.1769 [hep-th]].



\bibitem{Cvetic:2007fp}
G.~Cvetic and I.~Kondrashuk,
``Further results for the two-loop Lcc vertex in the Landau gauge,''
JHEP \textbf{02} (2008), 023
[arXiv:hep-th/0703138 [hep-th]].


\bibitem{Davydychev:1996pb}
A.~I.~Davydychev, P.~Osland and O.~V.~Tarasov,
``Three gluon vertex in arbitrary gauge and dimension,''
Phys. Rev. D \textbf{54} (1996), 4087-4113
[erratum: Phys. Rev. D \textbf{59} (1999), 109901]
[arXiv:hep-ph/9605348 [hep-ph]].


\bibitem{Davydychev:1992xr}
A.~I.~Davydychev,
``Recursive algorithm of evaluating vertex type Feynman integrals,''
J. Phys. A \textbf{25} (1992), 5587-5596


\bibitem{Kondrashuk:2008ec}
I.~Kondrashuk and A.~Kotikov,
``Fourier transforms of UD integrals,'' in B. Gustafsson and A. Vasil’ev (Eds.),  
Analysis and Mathematical Physics, Birkh\"auser Book Series Trends in Mathematics,
Birkh\"auser, Basel, Switzerland, 2009, pp. 337-348 
[arXiv:0802.3468 [hep-th]].


\bibitem{Kondrashuk:2008xq}
I.~Kondrashuk and A.~Kotikov,
``Triangle UD integrals in the position space,''
JHEP \textbf{08} (2008), 106
[arXiv:0803.3420 [hep-th]].



\bibitem{Kondrashuk:2009us}
I.~Kondrashuk and A.~Vergara,
``Transformations of triangle ladder diagrams,''
JHEP \textbf{03} (2010), 051
[arXiv:0911.1979 [hep-th]].


\bibitem{Erdmenger:1996yc}
J.~Erdmenger and H.~Osborn,
``Conserved currents and the energy momentum tensor in conformally invariant theories for general dimensions,''
Nucl. Phys. B \textbf{483} (1997), 431-474
[arXiv:hep-th/9605009 [hep-th]].


\bibitem{Freedman:1998tz}
D.~Z.~Freedman, S.~D.~Mathur, A.~Matusis and L.~Rastelli,
``Correlation functions in the CFT(d) / AdS(d+1) correspondence,''
Nucl. Phys. B \textbf{546} (1999), 96-118
[arXiv:hep-th/9804058 [hep-th]].

\bibitem{Gonzalez:2018gch}
I.~Gonzalez, I.~Kondrashuk, E.~A.~Notte-Cuello and I.~Parra-Ferrada,
``Multi-fold contour integrals of certain ratios of Euler gamma functions from Feynman diagrams: orthogonality of triangles,''
Anal. Math. Phys. \textbf{8} (2018) no.4, 589-602
[arXiv:1808.08337 [math-ph]].


\bibitem{Usyukina:1992jd}
N.~I.~Usyukina and A.~I.~Davydychev,
``An Approach to the evaluation of three and four point ladder diagrams,''
Phys. Lett. B \textbf{298} (1993), 363-370


\bibitem{Allendes:2012mr}
P.~Allendes, B.~Kniehl, I.~Kondrashuk, E.~A.~Notte-Cuello and M.~Rojas-Medar,
``Solution to Bethe-Salpeter equation via Mellin-Barnes transform,''
Nucl. Phys. B \textbf{870} (2013), 243-277
[arXiv:1205.6257 [hep-th]].

\bibitem{Borja:2016sap}
J.~Borja and I.~Kondrashuk,
``Alternative method of Reduction of the Feynman Diagrams to a set of Master Integrals,''
J. Phys. Conf. Ser. \textbf{762} (2016) no.1, 012056
[arXiv:1604.01353 [hep-th]].



\bibitem{Arkani-Hamed:2008owk}
N.~Arkani-Hamed, F.~Cachazo and J.~Kaplan,
``What is the Simplest Quantum Field Theory?,''
JHEP \textbf{09} (2010), 016
[arXiv:0808.1446 [hep-th]].





\end{thebibliography}
\end{document}